\documentclass[vecphys,pdftex]{svmult}

\pdfoutput=1
\usepackage{makeidx}         
\usepackage{graphicx}        
\usepackage[bottom]{footmisc}

\usepackage{bm}
\newcommand{\be}{\begin{equation}}
\newcommand{\ee}{\end{equation}}

\begin{document}

\title*{Spacetime: Arena or Reality?}
\author{H. I. Arcos\inst{1} \and
J. G. Pereira\inst{2}}
\institute{Universidad Tecnol\'ogica de Pereira, A.A. 97, La Julita, Pereira, Colombia \texttt{hiarcosv@utp.edu.co}
\and Instituto de F\'{\i}sica Te\'orica, Universidade Estadual Paulista, Rua Pamplona 145, 01405-900, S\~ao Paulo SP, Brazil \texttt{jpereira@ift.unesp.br}}

\maketitle

\section{Introduction}

The concept of fundamental particle has been quite elusive along the history of physics. The term {\it fundamental} is commonly used as a synonymous of structureless particles. However, this assumption is clearly contradictory. For example, it is impossible to explain spin without assuming a structure for the particle. In fact, a point particle is by definition spherically symmetric, a symmetry violated by the presence of spin. This problem is usually circumvented by saying that spin is a purely quantum property, which cannot be explained by classical physics. This means to keep it as a mysterious property of nature.

If one assumes that a fundamental particle is a point-like object, several arguments against this idea show up immediately. First, as discussed above, a point-like object seems to be inconsistent with the existence of spin. Second, if we try to reconcile general relativity with point-particles, which are singular points in a pseudo-Riemannian spacetime, unwanted features, like for example ultraviolet divergences, will appear. A natural alternative would be to assume that a fundamental particle is a string-like object, a point of view adopted by string theory \cite{gsw}. Similarly, one can introduce membranes as fundamental objects, or even extended objects with certain geometries. These models, however, are also plagued by problems. The membrane model has failed to generate a theory free of negatively-normed states, or tachyons, and theories with extended objects have failed to explain the existence of supporting internal forces that avoid the collapsing of the model.

With the evolution of particle physics and gravitation, the idea that a fundamental particle should somehow be connected to spacetime began to emerge. This is the case, for example, of Wheeler's approach, which was based on the concept of spacetime foam. At the Planck scale, uncertainty in energy allows for large curvature values. At this energy, spacetime can undergo deep transformations, which modify the small scale topology of the continuum. This is where the ``foam" notion becomes important. Small regions of spacetime can join and/or separate giving rise to non-trivial topological structures. The simplest of these structures is the so called wormhole, a quite peculiar solution to Einstein's equation. It represents a topological structure that connects spacetime points separated by an arbitrary spatial distance. An interesting property of the wormhole solution is that it can trap an electric field. Since, for an asymptotic observer, a trapped electric field is undistinguishable from a charge distribution, Wheeler introduced the concept of ``charge without charge'' \cite{mtw}. However, as Wheeler himself stated, these Planckian wormholes could not be related to any particle model for several reasons: charge is not quantized, they are not stable, their mass/charge ratio is very different from that found in known particles, and half-integral spin cannot be defined for a simple wormhole solution. There was the option to interpret a particle as formed by a collective motion of wormholes, in the same way phonons behave as particles in a crystal lattice. None of these ideas were developed further.

The discovery of the Kerr-Newman (KN) solution \cite{kerr,newman,newman2} in the early sixties opened the door for new attempts to explore spacetime-rooted models for fundamental particles \cite{lopez,otro,israel,bur1}. In particular, using the Hawking and Ellis extended interpretation of the KN solution \cite{hellis}, as well as the Wheeler's  concept of ``charge without charge'', a new model has been put forward recently \cite{electron}. The purpose of this chapter is to present a glimpse on the characteristics of this model, as well as to analyze the consequences for the concept of spacetime. We begin by reviewing, in the next section, the main properties and the topological structure of the KN solution.

\section{Kerr-Newman Solution}

\subsection{The Kerr-Newman Metric}

The stationary axially-symmetric Kerr-Newman (KN) solution of Einstein's equations
was found by performing a complex transformation on the tetrad field for the
charged Schwarzschild (Reissner-Nordstr\"om) solution \cite{kerr,newman,newman2}.
For $m^2\geq a^2+q^2$, it represents a black hole with mass $m$, angular momentum
per unit mass $a$, and charge $q$ (we use units in which $\hbar = c = 1$). In the
so called Boyer-Lindquist coordinates $r,\theta,\phi$, the KN solution is written as
\be \label{metric}
ds^2=dt^2-\frac{\rho^2}{\Delta} \, dr^2-(r^2+a^2) \sin^2\theta \, d\phi^2 -
\rho^2 \, d\theta^2-\frac{R r}{\rho^2} \, (dt - a\sin^2\theta \, d\phi)^2,
\ee
where
\[
\rho^2=r^2+a^2\cos^2\theta, \quad \Delta=r^2-R r+a^2, \quad R=2m-q^2/r.
\]
This metric is invariant under the change $(t, a)\rightarrow(-t, -a)$. It is also invariant under
$(m, r)\rightarrow(-m, -r)$ and $q \rightarrow -q$. This black hole is believed to be the final stage of a very general stellar collapse, where the star is rotating and its net charge is different from zero.

The structure of the KN solution changes deeply for $m^2<a^2+q^2$. Due to the absence of a horizon, it does not represent a black hole, but a circular naked singularity in spacetime. In fact, it represents a singular disk of radius $a$, whose border is a true singularity in the sense that it cannot be removed by any coordinate transformation. This means that there is a true singularity at the border. However, the metric singularity at the interior points of the disk can be removed by introducing a specific interpretation of the KN solution, as described by Hawking and Ellis \cite{hellis}. In what follows we give a detailed description of the topological structure behind such interpretation.

\subsection{The Hawking-Ellis Extended Interpretation}

The lack of smoothness of the metric components across the enclosed disk can be remedied by considering the extended spacetime interpretation of Hawking and Ellis \cite{hellis}. The basic idea of this extension is to consider that our spacetime is connected to another one through the interior points of the disk. This extended solution does not necessarily implies that the dimensionality of spacetime is greater than four, but rather that the manifold volume is greater than expected. In other words, the disk surface (with the upper points considered different from the lower ones) is interpreted as a shared border between our spacetime, denoted by {\bf M}, and another similar one, denoted by {\bf M'}. 
\begin{figure}[h]
\begin{center}
\includegraphics[height=4cm,width=5.5cm]{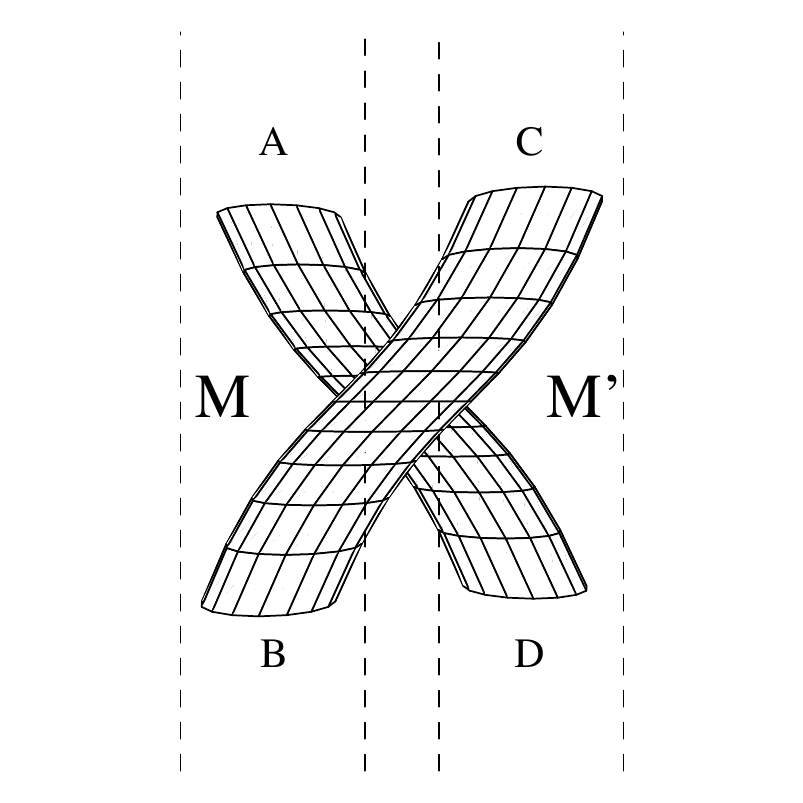}
\end{center}
\caption{To better visualize the intrinsic geometry of the KN manifold, the KN disk is drawn as if it presented a finite thickness, and consequently there is a space separation between the upper and lower surfaces of the disk. The left-hand side represents the upper and
lower surfaces of the disk in {\bf M}, whereas the right-hand side represents the
upper and lower surfaces of the disk in {\bf M'}.}
\label{graf1}
\end{figure}
According to this construction, the KN metric components are no longer singular across the disk, making it possible to smoothly join the two spacetimes, giving rise to a single $4$-dimensional spacetime, denoted ${\mathcal M}$. This link can be seen in Fig.~\ref{graf1} as solid cylinders going from {\bf M} to {\bf M'}. In this figure, to clearly distinguish the upper from the lower side, the disk was drawn as if it presented a finite thickness. In order to cross the disk, therefore, an electric field line that hits the surface A will forcibly emerge from surface D, in {\bf M'}. Then, it must go through surface C to finally emerge from surface B, in {\bf M}. This picture gives a clear idea of the topological structure underlying the KN solution.

Now, the singular disk is located at $\theta = \pi/2$ and $r=0$. Therefore, if $r$ is assumed to be positive in {\bf M}, it will be negative in {\bf M'}. Since the KN metric must be the same on both sides of the solution, the mass $m$ will be negative in {\bf M'}. Furthermore, the magnitude of the electric charge $q$ on both sides of the solution is, of course, the same. Taking into account that the source of the KN solution is represented by the electromagnetic potential
\be \label{potential}
{\bm A}= -\frac{qr}{\rho^2}(dt - a \sin^2\theta d\phi),
\ee
which is clearly singular along the ring, and since $r$ has different sign on different sides of the solution, we see from this expression that, if the charge is positive in one side, it must be negative in the other side.

\subsection{Causality versus Singularity}

As already remarked, the above extended interpretation does not eliminate the singularity at the rim of the disk. However, there are some arguments that can be used to circumvent this problem. First, it is important to observe that there is a torus-like region around the singular ring, in which the coordinate $\phi$ becomes timelike. Inside this region, defined by
\be\label{inequa}
r^2 + a^2+\left(\frac{rR}{\rho^2}\right) a^2\sin^2\theta < 0,
\ee
there will exist closed timelike curves \cite{carter}. In fact, when crossing the surface of this region, the signature of the metric changes from $(-,-,-,+)$ to $(-,-,+,+)$. This reduction in the number of spatial dimensions is a drawback of the solution.

\begin{figure}
\begin{center}
\includegraphics[height=2.4cm,width=9.8cm]{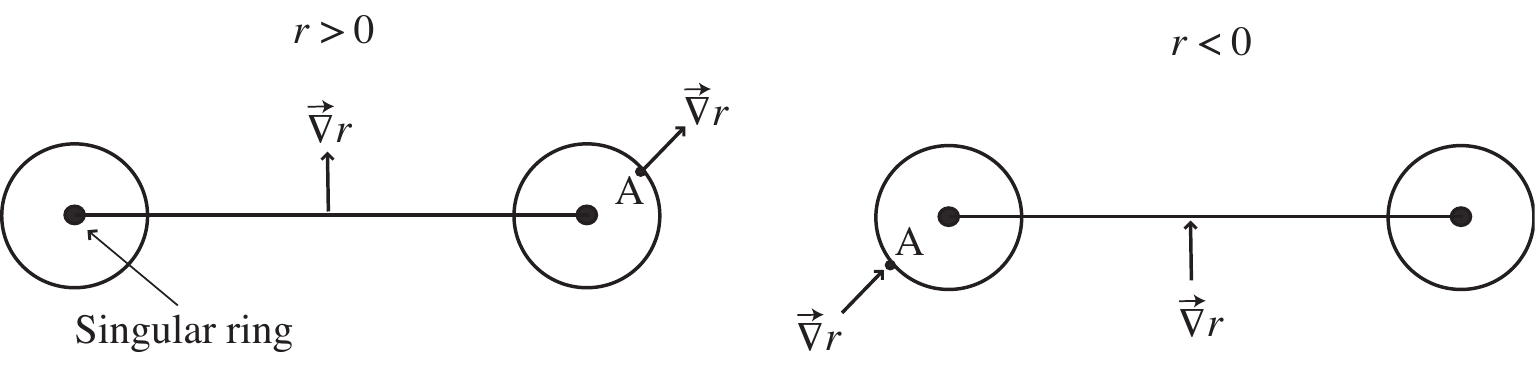}
\end{center}
\caption{Tubular-like regions around the singular ring, which is to be excised.
Several $\vec\nabla r$ directions are also depicted, which show how the borders in
the positive and negative $r$ sides can be continuously glued.}
\label{graf2}
\end{figure}

Now, when the values of $a$, $q$ and $m$ are chosen to be those of the electron, the surface of the torus-like region is separated from the singular ring by a distance of the order of $10^{-34}$ cm, which coincides roughly with the Planck length. At this scale, as is well known, topology changes are expected to exist, and consequently changes in the connectedness of spacetime topology are likely to occur. A solution to this problem is to excise the infinitesimal region around the singular ring on both the positive and negative $r$ sides, and then glue back the manifold.\footnote{This kind of singularity removal has already been explored by Punsly for the case of the Kerr solution \cite{punsly}.} A simple drawing of the region to be excised can be seen in Fig.~\ref{graf2}, where the direction of the gradient of $r$ has been drawn at several points.
As an example, note that the point $A$ on the positive $r$ side must be glued to the point $A$ on the negative $r$ side. If we glue all points of the torus border, we obtain a continuous path for the electric field lines that flow through the disk, even for those lines that would hit the disk at the singular ring. Furthermore, since the extrinsic curvature does not change sign when crossing the hypersurface $g_{\phi\phi}=0$, the above gluing process does not generate stress-energy \cite{israel}.

An important point of the above structure is that, after removing the tubular region around the singular ring, the surface delimiting both spacetimes turns out to be defined by a reversed topological product between two 2-torus. As is well known, this is nothing, but the Klein bottle \cite{carmo}. This is a crucial property because, as we are going to see, in order to present a spinorial behavior, any spacetime topological structure must somehow involve the Klein bottle. And of course, in order to be used as a model for any fundamental particle, a topological structure must necessarily be a spacetime spinorial structure. 
\index{Klein bottle}
\section{The KN Solution as a Dirac Particle}

\subsection{Preliminaries}

We are going now to explore the possibility of using the KN solution as a model for the electron. To begin with, let us observe that the total internal angular momentum $L$ of the KN solution, on either side of $\mathcal M$, can be written as
\be \label{cons3}
L = m \, a.
\ee
If we take for $a$, $m$ and $q$ the experimentally known electron values, and considering that, for a spin $1/2$ particle $L = 1/2$, it is easy to see that the disk has a diameter
equal to the Compton wavelength $\lambda/ 2\pi = 1/m$ of the electron. Consequently, the angular velocity $\omega$ of a point in the singular ring turns
out to be
\be
\omega = 2 \, m,
\ee
which corresponds to the so called {\it Zitterbewegung} frequency \cite{barut1,barut2} for a point-like electron orbiting a ring of diameter equal to $\lambda_e$. This means that the KN solution has a gyromagnetic ratio $g=2$ \cite{newman,carter}. Due to this property, several attempts to model the electron by using the KN solution have been made. In most of these models, however, the circular singularity was always surrounded by a massive ellipsoidal shell (bubble), so that it was actually unreachable. In other words, the singularity was considered to be non-physical in the sense that the presence of the massive bubble would preclude its formation.

Using the extended interpretation of Hawking and Ellis, a different model has been proposed recently \cite{electron}. Its main property is that, differently from older models, it is represented by an empty KN solution, that is, no surrounding massive bubble is supposed to exist around the singular ring.\footnote{A similar approach has been used by Burinskii; see \cite{bur2}, and references therein.} Instead, we make use of the excision procedure to circumvent the problems related to the naked singularity and the non-causal regions. The fundamental property of this model is that Wheeler's idea of ``charge without charge'' and ``mass without mass'' can be extended to spin. As a consequence, it is able to provide a topological explanation for the concepts of charge, mass, and spin.
\index{Charge without charge} \index{Mass without mass}

Charge can be interpreted as arising from the multi-connectedness of the spatial section of the KN solution. In other words, we can associate the electric charge of the KN solution with the net flux of a topologically trapped electric field. In fact, remember that, from the point of view of an asymptotic observer, a trapped electric field is indistinguishable from the presence of a charge distribution. Then, in analogy with the geometry of the wormhole solution, there must exist a continuous path for each electric field line going from one space to the other. Furthermore, the equality of magnetic moment on both sides of $\mathcal M$ implies that the magnetic field lines must also be  continuous when passing through the disk enclosed by the singularity.

Mass can be associated with the degree of non-flatness of the KN solution. It is given by Komar's integral \cite{komar},
\be \label{kint}
m = \int_{\partial\Sigma}\star d\xi,
\ee
which holds for any stationary, asymptotically flat spacetime. In this expression, $\star$ denotes the Hodge dual operator, $\xi$ is the stationary Killing one-form of the background metric, and $\partial\Sigma$ is a spacelike surface of the background metric. It should be noticed that the mass $m$ is the {\em total} mass of the system, that is, the mass-energy contributed by the gravitational and the electromagnetic fields \cite{ohanian}.

Finally, spin can be consistently interpreted as an internal rotational motion of the singular ring. Of course, after the excision process, it turns out to be interpreted as an internal rotation of the infinitesimally-sized Klein bottle. It is important to remark that the KN solution is a singular ring in spacetime, not in the three-dimensional space. In fact, if the singularity were, let us say, in the $xy$ plane, the angular momentum would be just a component of the {\it orbital} angular momentum, for which the gyromagnetic factor is well known to be $g=1$. Since the gyromagnetic factor of the KN solution is $g=2$, the rotation plane must necessarily involve the time axis. In fact, we know from Noether's theorem that conservation of {\it spin} angular momentum is related to the invariance of the system under a rotation in a plane involving the time axis.

\subsection{Wave-particle duality}

If one tries to compute the size of the KN particle, a remarkable result is obtained. To see it, we write down the spatial metric of the KN solution, which is given by \cite{landau}
\be\label{3metric}
dl^2 =\rho^2 \left[ \frac{1}{\Delta}dr^2 + d\theta^2 + \frac{\Delta\sin^2\theta}{\Delta - a^2\sin^2\theta}d\phi^2 \right].
\ee
If we use this metric to compute the spatial length ${\mathcal L}$ of the singular ring, we find it to be zero: 
\be
{\mathcal L} \equiv \int_0^{2\pi} dl = 0.
\ee
This result is consistent with previous analysis made by some authors \cite{israel,carter}, who pointed out that an external observer is unable to ``see" the KN solution as an extended object, but only as a point-like object. We can then say that the ``particle" concept is validated in the sense that the non-trivial KN structure is seen, by all observers, as a point-like object. Although the spatial dimension of the disk is zero, its spacetime dimension is of the order of the Compton wavelength for the particle, which for the electron is $\lambda = 10^{-11}$ cm.

It is well known that a fundamental particle fulfills the de Broglie relationship
\be \label{duality}
\lambda = \frac{1}{p}=\frac{1}{mv},
\ee
where $p$ is its total linear momentum. This relationship can be given a theoretical fundamentation in our model. The wave-lenght $\lambda$ is associated with the diameter of the singular ring, and at first glance it seems to be unrelated to mass. But since the radius $a$ is also the angular momentum per unit mass, which is a particular property of the solution, the mass $m$ and $a$ are linked by $ma=1/2$.

\subsection{Topological Structure}

\begin{figure}
\begin{center}
\includegraphics[height=5.5cm,width=6.5cm]{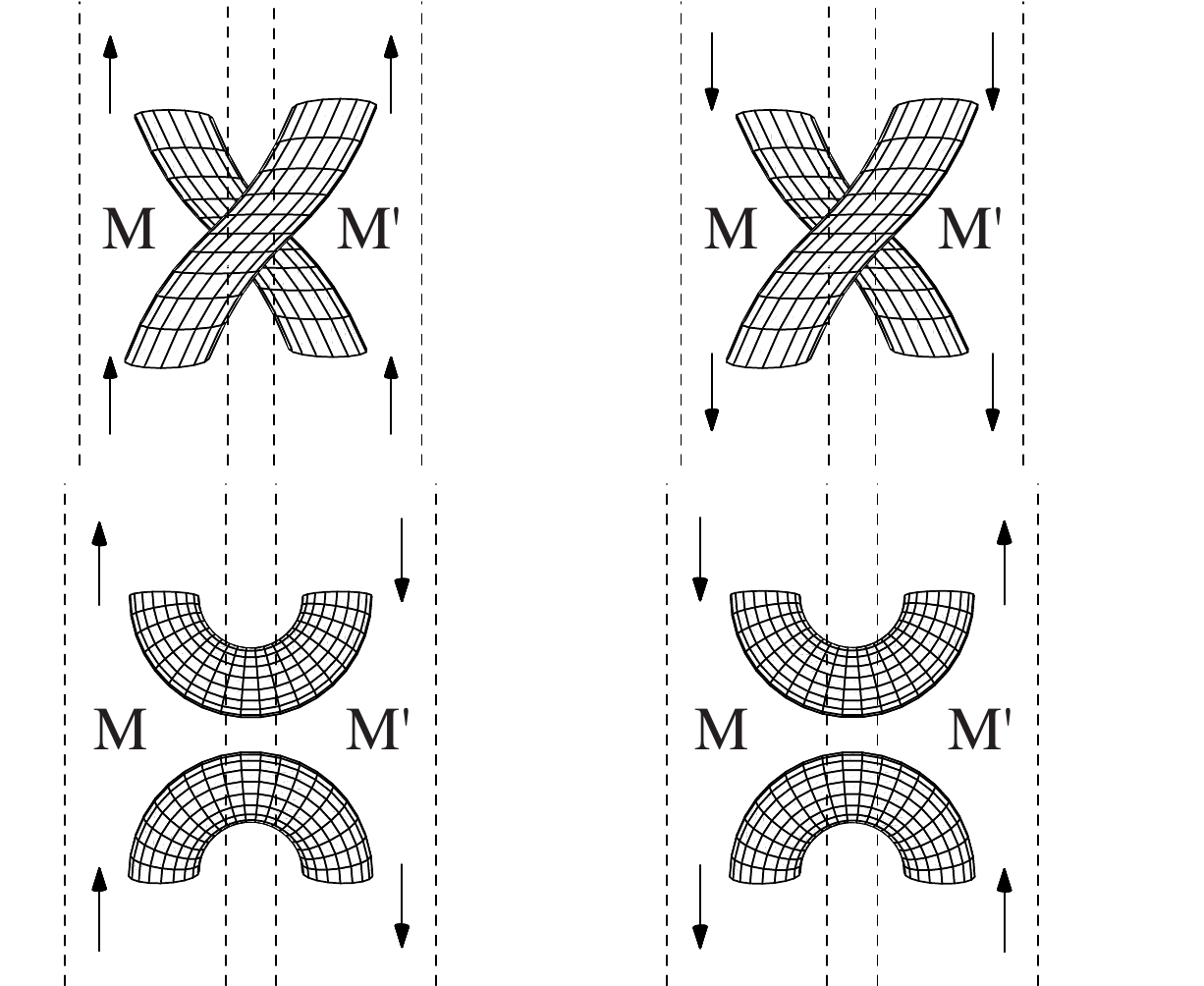}
\end{center}
\caption{The four possible geometric configurations of KN states for a specific value of the electric charge. The arrows indicate the sense of the spin vector.}
\label{graf3}
\end{figure}

A simple analysis of the structure of the extended KN metric shows that it is possible to
isolate {\em four} physically non-equivalent states on each side of ${\mathcal M}$,
that is, on {\bf M} and on {\bf M'}. These states can be labeled by the sense of
rotation ($a$ can be positive or negative), and by the sign of the electric charge
(positive or negative). Each one of these non-equivalent states in {\bf M} must be joined continuously through the KN disk to another one in {\bf M'}, but with opposite charge. Since we want a continuous joining of the metric components, this matching must take into account
the sense of rotation of the rings. In Fig.~\ref{graf3}, just as in Fig.~\ref{graf1}, the tubular joining between {\bf M} and {\bf M'} are drawn for one specific value of the electric charge,\footnote{Two signs for the electric charge $q$ in {\bf M} or {\bf M'} are allowed since the KN metric depends quadratically on $q$.} but taking into account the different spin directions in each disk, which are drawn as small arrows. The differences among the configurations are the orientation of the spin vector and the geometry of the tubes.

It is important to remark that the model considers both sides of the solution, that is, {\bf M} and {\bf M'}, as part of a single spacetime.\footnote{This is similar to the wormhole solution, which connects two points of the same spacetime.} The use of two spacetimes is just a mathematical necessity to describe the topological structure behind the KN solution. The question then arises on how to interpret the fact that the mass, and consequently the energy, acquires a negative value in {\bf M'}, if they are assumed to be positive in {\bf M}. The same happens with the sense of rotation, or equivalently, with the arrow of time. At this point it is possible to see the close analogy that exists between the topological structure of the KN solution and the structure of a Dirac spinor. In fact, the same questions on the interpretation of {\bf M} and {\bf M'} could be made on the interpretation of the two upper and the two lower components of the Dirac spinor. The answer to the latter question, as is well known, requires both special relativity and quantum mechanics, and consequently the notion of anti-particles to comply with negative energies \cite{weinberg}. We can then say that the necessity of two spacetimes to describe a spinorial structure in spacetime is quite similar to the necessity of a four-component spinor to describe a spin-half particle. 

\subsection{Existence of Spacetime Spinorial Structures}

The excision process used to eliminate the non-causal region gives rise to highly non-trivial topological structure. Now, it is a well known result that, in order to exhibit gravitational states with half-integral angular momentum, a 3-manifold must fulfill certain topological conditions. These conditions were stated by Friedman and Sorkin \cite{fs}, whose results were obtained from a previous work by Hendricks \cite{hendricks} on the obstruction theory in three dimensions. Interesting enough, the KN solution can be shown to satisfy these conditions, which means that it is actually a spacetime spinorial structure \cite{electron}.

An alternative way to verify this result is to analyze the behavior of the KN topological structure under rotations. In general, when rotated by $2\pi$, a classical object returns to its initial orientation. However, the topological structure of the KN solution presents a different behavior: it returns to its initial position only after a $4 \pi$ rotation. This result can be understood from the topology of the 2-dimensional surface that is formed in the excision and gluing procedure. This surface, as we have already seen, is just a Klein bottle. A $2\pi$ rotation of the positive $r$ side is equivalent to moving a point on the Klein bottle surface halfway from its initial position. Only after a $4\pi$ rotation it returns to its departure point. This is a well known property of M\"obius strip, and consequently of the Klein bottle since the latter is obtained by a topological product of two M\"obius strips.

\subsection{Evolution Equation}

As we have seen, the extended KN solution represents a spacetime spinorial structure. It can, therefore, be naturally represented in terms of spinor variables of the Lorentz group SL(2, $\mathbf{C}$). A crucial point towards this possibility is the fact that the KN solution presents four non-equivalent states, defined by the sense of rotation and by the sign of the electric charge. Since a Dirac spinor also has four independent components, it is not difficult to find an algebraic representation for the KN solution. Considering then an asymptotic observer in a Lorentz frame moving with a constant velocity, the evolution of the KN state vector is found to be governed by the Dirac equation \cite{electron}. Taking into account that the KN solution represents a spacetime spinorial structures, we can say this is a natural and expected result.

\section{Concluding Remarks}

By using the extended spacetime interpretation of Hawking and Ellis, together with Wheeler's idea of ``charge without charge'', the KN solution was shown to exhibit properties that are quite similar to those presented by an electron. Apart from the eventual importance of this result for particle physics, there is also deep consequence for the concept of spacetime. At the early times of gravitation theory, space was considered simply an arena where all phenomena would take place. In other words, space was just a relation between the existing objects; without objects, there would be no space. Later on, the existence of an aether was considered, which in a sense would give some reality to the space. Since all experiments to detect such aether gave null results, space continued for some time to be this {\it mysterious nothing} in which we live in.

The advent of special relativity introduced the first important changes in our concept of space. Time lost its absolute character, and became just one more coordinate. Instead of living in a three-dimensional space, we discovered that we actually live in a four-dimensional spacetime. The advent of general relativity introduced further and deeper conceptual changes in our notion of spacetime. We discovered, for example, that spacetime can storage energy. This means essentially that it could not anymore be interpreted as a simple arena because, if it can storage energy, it must have a concrete existence.

In addition to simple configurations, like a curved spacetime, general relativity allows the existence of much more complex spacetime structures. One example is the KN solution of Einstein's equation, which presents a very peculiar topological structure. Its main property is to be a spinorial spacetime structure, which is revealed by the fact that only after a $4 \pi$ rotation it returns to its initial position. The presence of the Klein bottle in the topological structure makes it easier to understand this property.

Now, if we consider that the topological structure is able to trap an electric field, an asymptotic observer would see it as if the structure presented an electric charge. Furthermore, because the curved spacetime associated to the topological structure has a non-vanishing energy, the same asymptotic observer would see it as if the structure presented a mass. When the experimental values for the electron charge and mass are used, the internal angular momentum of the KN solution is found to present a gyromagnetic factor $g = 2$. In addition to storage energy, therefore, spacetime can also carry electric charge and spin angular momentum.

Due to the fact that it represents a spacetime spinorial structure, the KN solution can be represented in terms of the spinor variables of the Lorentz group SL(2, $\mathbf{C}$). Its spacetime evolution is then naturally found to be governed by the Dirac equation. The KN structure, therefore, can be interpreted as a spacetime-rooted electron model. Of course, it is not a finished model, and many points remain to be understood and clarified. For example, it is an open question whether it is applicable or not to other particles of nature. If, however, it shows to be a viable model, spacetime will acquire a new and more important status. In fact, it will be not only the arena, but will also provide --- through its highly non-trivial Planck-scale topological structures --- the building blocks of all existing matter in the universe, including ourselves.

\section*{Acknowledgments}
The authors would like to thank A. Burinskii and T. Nieuwenhuizen for useful comments. They would like to thank also FAPESP, CNPq and CAPES for financial support.



\begin{thebibliography}{99.}

\bibitem{gsw}
M. B. Green, J. H. Schwarz and E. Witten: \textit{Superstring theory} (Cambridge University Press, Cambridge, 1988).

\bibitem{mtw}
J. A. Wheeler: {\it Geometrodynamics} (Academic Press, New York, 1962).

\bibitem{kerr}
R. P. Kerr: {Phys. Rev. Lett.} \textbf{11}, 237 (1963).

\bibitem{newman}
E. T. Newman and A. I. Janis: {J. Math. Phys.} \textbf{6}, 915 (1965).

\bibitem{newman2}
E. T. Newman {\it et al}: {J. Math. Phys.} \textbf{6}, 918 (1965).

\bibitem{lopez}
C. A. Lopez: {Phys. Rev.} \textbf{D30}, 313 (1984); C. A. Lopez: {Gen. Rel. Grav.} \textbf{24}, 285 (1992).

\bibitem{otro}
M. Israelit and N. Rosen: {Gen. Rel. Grav.} \textbf{27}, 153 (1995).

\bibitem{israel}
W. Israel: {Phys. Rev.} \textbf{D2}, 641 (1970).

\bibitem{bur1}
A. Burinskii: {Sov. Phys. JETP} \textbf{39}, 193 (1974).

\bibitem{hellis}
S. W. Hawking and G. F. R. Ellis: {\it The Large Scale Structure of
Space-Time} (Cambridge University Press, Cambridge, 1973) p. 161.

\bibitem{electron}
H. I. Arcos and J. G. Pereira: Gen. Rel. Grav. \textbf{36}, 2441 (2004) [hep-th/0210103]

\bibitem{carter}
B. Carter: {Phys. Rev.} \textbf{174}, 1559 (1968).

\bibitem{punsly}
B. Punsly: {J. Math. Phys.} \textbf{28}, 859 (1987).

\bibitem{carmo}
M. P. do Carmo: {\it Differential Geometry of Curves and Surfaces}
(Prentice-Hall, New Jersey, 1976).

\bibitem{barut1}
A. O. Barut and A. J. Bracken: {Phys. Rev.} \textbf{D23}, 2454 (1981).

\bibitem{barut2}
A. O. Barut and W. Thacker: {Phys. Rev.} \textbf{D31}, 1386 (1985).

\bibitem{bur2}
A. Burinskii: {Phys. Rev.} \textbf{D68}, 105004 (2003).

\bibitem{komar}
A. Komar: {Phys. Rev.} \textbf{113}, 934 (1959).

\bibitem{ohanian}
H. Ohanian and R. Ruffini: {\it Gravitation and Spacetime} (Norton \&
Company, New York, 1994) p. 396.

\bibitem{landau}
L. D. Landau and E. M. Lifshitz: {\it The Classical Theory of Fields}
(Pergamon, Oxford, 1975).

\bibitem{weinberg}
S. Weinberg: {\it Gravitation and Cosmology} (Wiley, New York, 1972), page 61.

\bibitem{fs}
J. L. Friedman and R. Sorkin: {Phys. Rev. Lett.} \textbf{44}, 1100 (1980).

\bibitem{hendricks}
H. Hendriks: {Bull. Soc. Math. France Memoire} \textbf{53}, 81 (1977); see section
4.3

\end{thebibliography}
\end{document}